# Data-driven modeling for boiling heat transfer: using deep neural networks and high-fidelity simulation results

Yang Liu[1,*], Nam Dinh[1], Yohei Sato[2], Bojan Niceno[2]
1 Department of Nuclear Engineering, North Carolina State University, USA
2 Paul Scherrer Institute, Switzerland

**Abstract**

Boiling heat transfer occurs in many situations and can be used for thermal management in various engineered systems with high energy density, from power electronics to heat exchangers in power plants and nuclear reactors. Essentially, boiling is a complex physical process that involves interactions between heating surface, liquid, and vapor. For engineering applications, the boiling heat transfer is usually predicted by empirical correlations or semi-empirical models, which has relatively large uncertainty. In this paper, a data-driven approach based on deep feedforward neural networks is studied. The proposed networks use near wall local features to predict the boiling heat transfer. The inputs of networks include the local momentum and energy convective transport, pressure gradients, turbulent viscosity, and surface information. The outputs of the networks are the quantities of interest of a typical boiling system, including heat transfer components, wall superheat, and near wall void fraction. The networks are trained by the high-fidelity data processed from first principle simulation of pool boiling under varying input heat fluxes. State-of-the-art algorithms are applied to prevent the overfitting issue when training the deep networks. The trained networks are tested in interpolation cases and extrapolation cases which both demonstrate good agreement with the original high-fidelity simulation results.

## 1. Introduction

Nucleate boiling, including pool boiling and subcooled flow boiling, is a highly efficient heat transfer regime, which is used for thermal management in various engineered systems with high energy density, from power electronics to heat exchangers in power plants and nuclear reactors. Essentially, nucleate boiling is a complex physical process that involves interactions among heating surface, liquid, and vapor, including nucleation, convection, evaporation, and surface properties. The mechanisms underlying such processes are still not fully understood, thus making accurate prediction of the heat transfer behavior under nucleate boiling regime a challenging task.

---

[*] Corresponding author.
Email address: yliu73@ncsu.edu (Y. Liu), ntdinh@ncsu.edu (N. Dinh), yohei.sato@psi.ch (Y. Sato), bojan.niceno@psi.ch (B. Niceno)

For engineering applications related to boiling, the major interest is to predict the heat transfer behavior of the two phase and boiling system, the quantities of interests (QoIs) often include the wall superheat, the heat transfer coefficient, as well as the evaporation source. A classical approach in nuclear technology for prediction of those QoIs are the system codes such as TRACE [49] which rely on empirical correlations, making their application less general. In more recent practices, the two-fluid model based multiphase computational fluid dynamics(MCFD) solver has been widely regarded as a promising tool for dealing with the boiling scenario, especially for systems with complex geometries such as nuclear reactor's fuel rod bundle. One of the major advantages of the two-fluid-model is that it averages the interface information between vapor and liquid with the help of closure models, thus significantly reducing the computational requirements compared to the first-principle simulation methods such as volume of fluid (VOF) or interface tracking method (ITM). On the other hand, however, the two-fluid model requires closures to be introduced to make the conservation equations for individual phases solvable. For example, the wall boiling closure is necessary in two-fluid-model to predict wall boiling behavior. Such closure consists of a combination of mechanistic models and empirical correlations, including nucleation site density, bubble departure diameter/frequency and heat partitioning. Great efforts have been put for the development of those models and correlations. Representative works on the study of nucleation include Cole [9], Kocamustafaogullari [24], Wang and Dhir [50], Hibiki and Ishii [19]. Besides that, the heat partitioning methods are also actively studied [4, 14, 22, 26]. Applications based on MCFD solver have been conducted and promising results have been obtained [13, 17, 34]. Yet a comprehensive review by Cheung et al. [7] observed that no single combination of these closure relations was able to provide satisfactory predictions for all QoIs over a variety of input conditions. Such limitations can be explained by two types of uncertainty inherent in the models.

One is the **model parameter uncertainty**. A model always contains parameters that need to be specified before it can be used for prediction. If a certain phenomenon is not well understood, the model describing it would inevitably contain empirical parameters that do not have clear physical interpretations and whose values are developed only to fit certain experimental measurements. The model parameter uncertainty mainly stems from these empirical parameters. The other is the **model form uncertainty,** which is embedded in the fixed functional form of the model. For a phenomenon which is not well understood, the formulation of the model for that phenomenon usually has to include approximations and simplifications. Thus, even knowing the true values of all the parameters, a model developed in this way still cannot describe the relevant physics with enough accuracy. To address these two sources of uncertainty of the traditional models, the uncertainty quantification (UQ) with sensitivity analysis (SA) is an approach with several applications already [20, 38, 52].

Meanwhile, it is worth to notice that the machine learning method provides insights in addressing the uncertainty issues and thus can be used to improve the predictive capability of the model. For example, the artificial neural network (ANN) has been regarded as a helpful machine learning tool for some complicated physical problems. Scalabrin et al. [44] apply ANN to study the heat transfer behavior in boiling process. Chang et al.[6] use ANN to predict heat transfer in supercritical water. Singh and Abbassi[46] couple CFD and ANN to model the transient behavior of Heating, ventilation and air conditioning (HVAC) systems. Parrales et al.[36] use ANN to model the void fraction for two-phase flow in helical heat exchangers. Alimoradi and Shams[1] use ANN and genetic algorithms to optimize the subcooled flow boiling in a vertical pipe. Prieler et al. (Prieler et al., 2018) use the similar methods to optimize the transient heating process in a natural gas fired furnace. Azizi et al.[2] studies the water hold-up of two phase flow in vertical and inclined tubes using ANN. Ma et al. [32] use neural networks learn from DNS data to construct closures to represent the interfacial force in MCFD solver. Hassanpour et al.[18] test different ANNs to predict the pool boiling heat transfer coefficient of alumina water-based nanofluids. It is observed that for all these applications, the applied ANNs have relative simple architecture with only one or two hidden layers and use global information such as mass flow rate, vapor quality. as input features. In turn, the obtained outputs from the ANNs are also macro-scale physical quantities, such as flow regime, cross-sectional averaged void fraction, or a universal heat transfer coefficient over the applicable domain. Such works demonstrates their applicability on system level code and serve as an improved version of an empirical correlation.

On the other hand, with the development of algorithms and computational power, the deep neural network (DNN), a version of ANN with more hidden layers and more sophisticated architectures, starts demonstrating its power and draws increasing attention with a series of successful applications on several topics [28]. Recently, there are wide applications of DNN on several areas including computer vision [25], natural language processing [21], and playing the sophisticated Go game [45]. The DNNs have some good mathematical properties for dealing with complex problems. It has been proven by Barron [3] that even a two-layer ANN of sigmoid activation function can approximate any form of continuous function, given large enough hidden units and properly chosen weights. It is further proven by Eldan and Shamir [11] that a DNN can achieve better **expressiveness** with more flexibility. Such property of DNN indicates a properly trained DNN can avoid the model form uncertainty. Moreover, DNN has good **optimization** property. It is also observed from practices that a DNN can be optimized and converges towards global minimum with a straightforward method, i.e. the stochastic gradient descent (SGD), although the theoretical explanation of its success is still weak [30]. The last good property is the **generalization** capability of DNN. It is observed through practices in natural language processing and computer vision, that a DNN usually has good performance even in predicting cases it has not been trained on [54]. On the other hand, although featuring many good properties, DNN usually requires large amount of data to train before it can be used for predictions. Moreover, the performance of DNN is highly dependent on the hyperparameters applied in the training process.

For physical problems, DNN also demonstrates its potential as a promising data-driven modeling approach. Compared to the aforementioned ANN applications, current DNN can capture local information with the support of local features, thus significantly expanding its potential application to the thermal fluid community. Li et al. [29] use deep belief network to model the post-combustion CO2 capture process. Another active research topic is the data-driven turbulence modeling. Ling et al. [31] use DNNs with the input of invariant tensor basis to predict the anisotropy tensor in a turbulent flow. The DNN is trained by single phase turbulent flow direct numerical simulation (DNS) results. The trained DNN demonstrated significant improvement compared to baseline Reynolds-averaged Navier-Stokes (RANS) eddy viscosity models. Inspired by this, Kutz [27] predicts the DNNs would play a significant role in turbulence modeling in next decade. Moreover, A comprehensive review of the application of DNN in thermal fluid related problems is performed by Chang and Dinh [5]. It should be noted from the review that, in the current stage, there are still only limited applications of DNNs on physical problems and such applications are more focused on turbulence modeling. To authors' best knowledge, there are still no application of DNN for boiling related problems. One of the obstacle to apply DNN to boiling related problems is the difficulty to obtained high-fidelity and high-resolution data that contains the local information of the boiling scenarios. Current experimental measurement technology still cannot provide accurate measurement on the detailed near wall flow and boiling features, especially for high heat flux scenario where the nucleate boiling occurs intensively. On the other hand, high-fidelity simulation of two phase flow with phase change provides an alternative high-fidelity data source for the DNN. Sato and Niceno [40, 41] perform first principle simulation with interface tracking on nucleate pool boiling under varying wall heating conditions, taking into account conjugate heat transfer between the wall and the fluid. The simulation results achieved good agreement with the experimental measurement and thus can be regarded as high-fidelity results. Data extracted from those simulation results are used for training the DNN studied in this paper.

It is also worthwhile to mention that besides the neural networks, some other machine learning methods have also attracted increased attention in the study of thermal fluid related problems. Parish and Duraisamy [35] use Bayesian method to inversely infer field variables as an approach to address the model form uncertainty. Wang et al. [51] use random forest to represent the discrepancy function between results from RANS equations and DNS results, the trained discrepancy function can be applied to difference scenarios. One common feature of all those aforementioned methods is the focus on addressing the model form uncertainty, whether by introducing a new variable representing it (such as the field variable or error estimate function) or by developing a novel form of closures with enough flexible representations to avoid it (such as the DNNs).

In this paper, a study is performed which aims to apply the DNNs to predict the wall heat transfer behavior in the context of two fluid models. The local flow feature, surface feature and heating feature are chosen as the inputs of the DNNs. The aforementioned ITM simulation results on conjugate pool boiling under quasi-steady state are extracted and averaged under the Eulerian framework. The obtained data are served as training and validation dataset. The QoIs of the system, including near wall void fraction, wall superheat, and two heat transfer components are the outputs of the DNNs. The paper is organized in the following structure. Section.2 discusses the fundamentals of DNN. Section.3 introduces the data processing procedure on ITM simulation results. Section.4 describes the problem setup, including neural network architecture and the data extraction work. Section.5 discusses the obtained results. Section.6 gives the summary remarks of the work.

**2. Deep feedforward networks**

This work is based on the deep feedforward neural networks (DFNNs), which is the most fundamental type of DNN in practical applications, as illustrated in Figure 1. In deep feedforward network, the first layer is the input and the last layer is the output, in between are the hidden layers, the units in hidden layer are term "neurons" and are represented by a nonlinear function of certain form. The term "deep" indicates the feedforward network has multiple hidden layers, while the term "feedforward" means there are no feedback connections in the hidden layers.

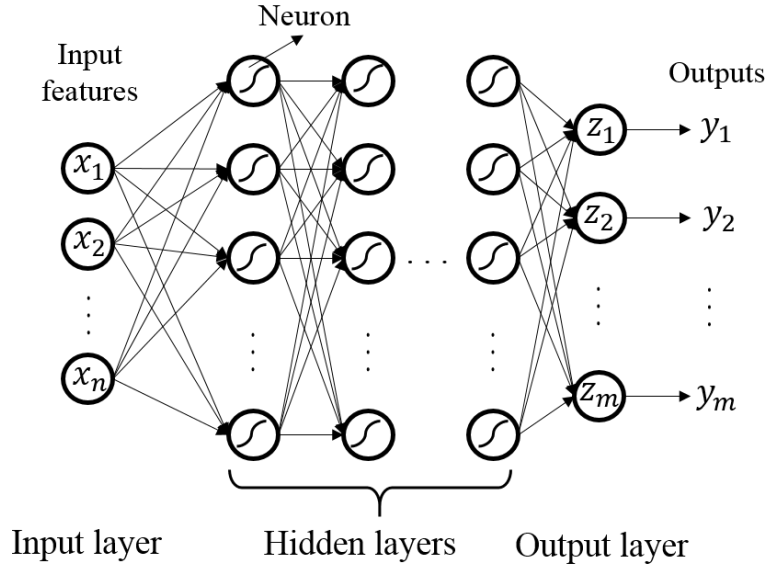

**Figure 1. Architecture of a fully connected deep feedforward network.**

Essentially, the DFNN can be regarded as a process where the input features go through a series of nonlinear transformations to predict the outputs as QoIs. Most DFNNs do so using an affine transformation controlled by learnable parameters weights $W$ and biases $b$, followed by a nonlinear function named activation function $g(x)$. In this sense, the neural network in Figure 1 can be interpreted as the following transformations:

$$\begin{aligned} \boldsymbol{h}_1 &= g(\boldsymbol{W}_1^T \boldsymbol{x} + \boldsymbol{b}_1) \\ \boldsymbol{h}_2 &= g(\boldsymbol{W}_2^T \boldsymbol{h}_1 + \boldsymbol{b}_2) \\ &\vdots \\ \boldsymbol{y} &= g(\boldsymbol{W}_{l+1}^T \boldsymbol{h}_l + \boldsymbol{b}_{l+1}) \end{aligned} \quad (1)$$

There are many choices of activation function in practice, including sigmoid function, tanh function, rectified linear units (ReLU) [33]. In this work, a modified version of ReLU, the Exponential linear unit (ELU) [8] is chosen which can be expressed as follows:

$$g(x) = \begin{cases} \alpha(e^x - 1), & x < 0 \\ x, & x \geq 0 \end{cases} \quad (2)$$

With the input features, number of hidden layer, number of hidden units, weights $W$ and biases $b$, activation function, and outputs setup, the architecture of the DFNN is determined. It will accept an input feature vector $x$ and propagate through every hidden layer to produce an output vector $\hat{y}$, the weight matrix $W$ and bias vector $b$ determines the prediction. Without proper training, it cannot be guaranteed that $\hat{y}$ can approximate the real data $y$, thus the prediction would be meaningless. The next step is to train the DFNN with enough data, generally speaking, the deeper the network, the more data required for training. For training a DFNN, a loss function $L(\hat{y}, y)$ need to be defined to measure the error between the DFNN prediction $\hat{y}$ and the real data $y$. For physical problems, the $L_1$ and $L_2$ norm loss functions are most widely used:

$$\begin{aligned} L_1 \; norm: L(\hat{y}, y) &= \|\hat{y} - y\|_1 = \frac{1}{m}\sum_{i=1}^{m}|\hat{y}_i - y_i| \\ L_2 \; norm: L(\hat{y}, y) &= \|\hat{y} - y\|_2 = \frac{1}{m}\sum_{i=1}^{m}(\hat{y}_i - y_i)^2 \end{aligned} \quad (3)$$

Once the loss function is defined the error gradients with regard to the weights and biases of each layer can be computed through the backpropagation based on the chain rule of calculus. Thus, the training process is a repeat of two steps: first input the data, do a forward propagation through the network, then perform the back propagation to obtain the error derivatives with regard to the weights and biases of each layer. Figure 2 demonstrates such procedure with a simple two hidden layer units and $L_2$ norm loss function, the bias is assumed to be zero for simplicity. In the figure, the notation $w_{ij}$ is the weight component between neuron $i$ and neuron $j$. The detailed backpropagation algorithm proposed by Rumelhart et al. [39] is summarized in Appendix. A.

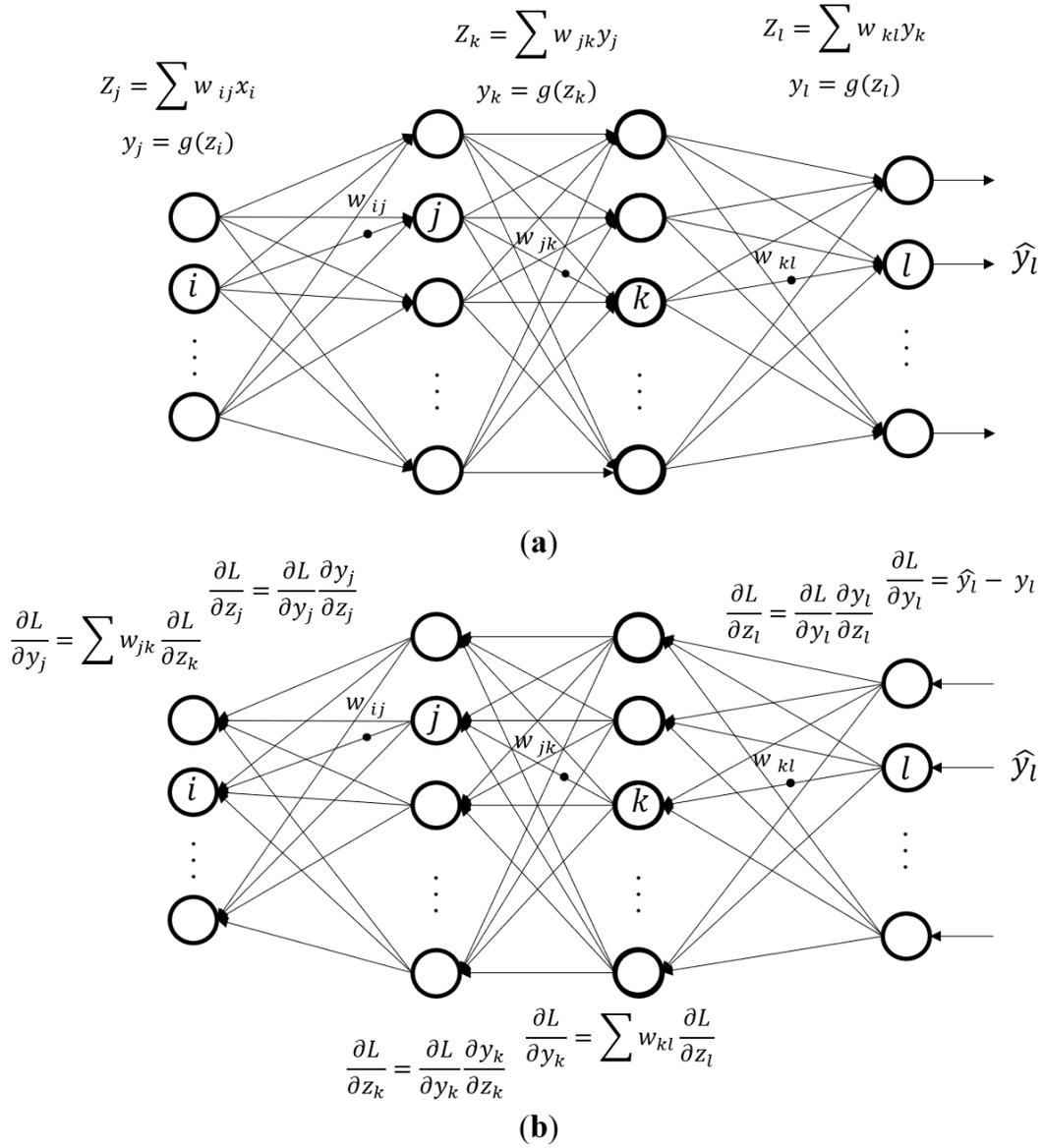

**Figure 2. Demonstration of (a) forward-propagation of input features; (b) back-propagation of loss function gradients.**

Based on the computed gradients, the weights and biases can be updated with the stochastic gradient descendent algorithm, the goal is to find a set of weights and biases (denoted by **θ** here) for every layer to minimize the loss of all data:

$$J(\boldsymbol{\theta}; \hat{\boldsymbol{y}}, \boldsymbol{y}) = \frac{1}{N} \sum_{i=1}^{N} L(\hat{\boldsymbol{y}}^{(i)}, \boldsymbol{y}^{(i)}, \boldsymbol{\theta}) \tag{4}$$

$$\nabla_{\boldsymbol{\theta}} J(\boldsymbol{\theta}; \widehat{\boldsymbol{y}}, \boldsymbol{y}) = \frac{1}{N} \sum_{i=1}^{N} \nabla_{\boldsymbol{\theta}} L(\widehat{\boldsymbol{y}}^{(i)}, \boldsymbol{y}^{(i)}, \boldsymbol{\theta}) \qquad (5)$$

$$\boldsymbol{\theta}_{update} = \boldsymbol{\theta} - \varepsilon \nabla_{\boldsymbol{\theta}} J(\boldsymbol{\theta}; \widehat{\boldsymbol{y}}, \boldsymbol{y}) \qquad (6)$$

In this process, $\varepsilon$ is termed learning rate which is the most important parameter during the training whose value directly affect the performance of the network. In practice, a more effective training process is to randomly sample many mini-batches from the full dataset and update $\boldsymbol{\theta}$ with every mini-batch. The size of the mini-batch is usually around hundred, level no matter how large the full dataset is. Moreover, more advanced algorithm improved from the standard SGD is usually applied in practice. In this work, the adaptive moment estimation (Adam) algorithm proposed by Kingma and Ba [23] is applied which computes individual adaptive learning rates for different parameters from estimates of first and second moments of the gradients. The training process would go over the datasets many iterations (termed epoch in the deep learning community).

In the practices of training a very deep DFNN, there are still two common issues: the overfitting and the potential vanishing or exploding gradients. If a DFNN is overfitted with the training data, the total loss function would be small, but it will fail to predict dataset that is not included in the training. A way to minimize the overfitting is to include a regularization term $\boldsymbol{\Omega}(\boldsymbol{\theta})$ in the loss function $L$:

$$\tilde{J}(\boldsymbol{\theta}; \widehat{\boldsymbol{y}}, \boldsymbol{y}) = J(\boldsymbol{\theta}; \widehat{\boldsymbol{y}}, \boldsymbol{y}) + \lambda \boldsymbol{\Omega}(\boldsymbol{\theta}) \qquad (7)$$

where $\lambda$ is a positive hyperparameter that weights the relative contribution of the regularization term, large $\lambda$ indicates strong regularization. In most practices, the regularization term is only applied to weights, thus $\boldsymbol{\Omega}(\boldsymbol{\theta})$ is equal to $\boldsymbol{\Omega}(\boldsymbol{w})$. Most widely used $\boldsymbol{\Omega}(\boldsymbol{w})$ are $L_1$ and $L_2$ regularization:

$$\begin{aligned} L_1 \ regularization: \quad & \boldsymbol{\Omega}(\boldsymbol{w}) = \|\boldsymbol{w}\|_1 = \sum w_i \\ L_2 \ regularization: \quad & \boldsymbol{\Omega}(\boldsymbol{w}) = \frac{1}{2} \boldsymbol{w}^T \boldsymbol{w} \end{aligned} \qquad (8)$$

In this work, the $L_2$ regularization is applied. Moreover, the total dataset is divided into the training data and test data. to exam if a DFNN is overfitted or not. The test data is not included in the training process but is used to test the accuracy of the trained DFNN for inputs that it did not know in the training.

When training a very deep DFNN, the gradients can sometimes get either very big or very small through the backpropagation process. Such unstable behavior makes training difficult or even fail to converge. Currently, there is no universal solution for such issue. A partial solution is to carefully chose the initialization of weights. In this work, the weights are initialized with state-of-the-art method, and are also closely monitored during the training process.

To summarize: although the DNN has demonstrate its power in many fields, there are still several unclear aspects about its property. In current practices, training a DNN depends on experience and is usually a trial-and-error process. The predictive capability of DNN is highly dependent on not only data but also its hyperparameters: the learning rate $\varepsilon$, the number of hidden layers, the number of hidden units in each layer, the regularization coefficient $\lambda$, the mini-batch size, etc. In this work, the effects of those hyperparameters are investigated.

**3. Data processing**

The data used for training DNN in this work is obtained from ITM simulation, which is based on directly solving the incompressible Navier-Stokes equations with a sharp-interface, phase-change model proposed in [42]. The three conservation equations solved in this approach can be expressed as follows:

Mass:

$$\frac{\partial \rho}{\partial t} + \nabla \cdot \rho \boldsymbol{U} = 0 \tag{9}$$

Momentum:

$$\frac{\partial \rho \boldsymbol{U}}{\partial t} + \nabla \cdot (\rho \boldsymbol{U}\boldsymbol{U}) = -\nabla p + \nabla \cdot \{\mu(\nabla \boldsymbol{U} + (\nabla \boldsymbol{U})^T)\} + \boldsymbol{f} \tag{10}$$

Energy:

$$C_p \left(\frac{\partial T}{\partial t} + \boldsymbol{U} \cdot \nabla T\right) = \nabla \cdot (\lambda \nabla T) + Q \tag{11}$$

In addition to the conservation equations, the color function $\phi$ is used to track the interface between vapor and liquid:

$$\frac{\partial \phi}{\partial t} + \nabla \cdot (\phi \boldsymbol{U}) = -\frac{1}{\rho_l} \Gamma_{lv} \tag{12}$$

In the simulation, the nucleation site is prescribed in the whole heating surface, and the heat conduction in the solid wall is considered through conjugate heat transfer. With the boundary conditions specified, the solver is able to predict the detailed boiling process with high accuracy. The QoIs of boiling process include the wall superheat, evaporation heat transfer component, convective heat transfer component towards liquid, and near wall bubble concentration. Such QoIs are the outcome of complex interactions between different phenomena, including: convection, evaporation, conjugate heat transfer, buoyancy, and nucleation. Although it is impossible to develop an explicit correlation or a model to accurately account for such interaction and give a reasonable prediction for the QoIs, a DNN that takes those phenomena as input features can serve as a "black-box" model for the prediction of boiling QoIs and can be applied to untested conditions.

On the other hand, however, the ITM simulation is performed on very fine meshes, the results contain detailed interface information, as well as the fluctuation of physical quantities. For the two-fluid-model, such information is not only unnecessary but also incompatible to the averaged conservation equations. In this sense, the obtained ITM results need to go through certain average process before being used for training the DNN which is supposed to be compatible with two-fluid-model.

In this work, a combination of time and space average is processed for each physical quantities $f(\boldsymbol{x}, t)$ into a space and time averaged form $\langle f \rangle(\boldsymbol{x}, t)$ which can be described as follows:

$$\langle f \rangle(\boldsymbol{x}, t) = \frac{1}{\tau} \frac{1}{l^3} \int_{t-\tau}^{t} \int_{x_1-l/2}^{x_1+l/2} \int_{x_2-l/2}^{x_2+l/2} \int_{x_3-l/2}^{x_3+l/2} f(\boldsymbol{x}', t') dx'_3 dx'_2 dx'_1 dt' \tag{13}$$

Where $\tau$ and $l$ are the averaging time scale and averaging length scale, respectively. One interesting fact that worth noting, as discussed by Drew [10], is that this average process is mathematically equivalent to the convolution operation over a 4-dimensional matrix (three dimension in space, one dimension in time), with a kernel function $g(\boldsymbol{x})$:

$$\langle f \rangle(\boldsymbol{x}, t) = \int\int\int\int_{R^4} g(\boldsymbol{x} - \boldsymbol{x}') f(\boldsymbol{x}', t) d\boldsymbol{x}' \tag{14}$$

For quantities that are only valid in the heating surface, i.e. the potential nucleation site density, nucleation activation temperature, wall superheat, and the heat transfer components, the average process is performed on two-dimensional surface and time. Such operation is widely adapted in the DNN for extracting and preserving features of the data. It is assumed the averaged process will preserve the causal-relationship between input features and the boiling QoIs. Such assumption is reasonable if the ITM simulation already reached quasi-steady state before the data is extracted. Every physical quantity obtained from the ITM simulation are propagated through the process to generate an averaged version of it. The void fraction $\alpha$ is obtained by applying color function $\phi$ through this process. One example of the average process is demonstrated in Figure 3, where the color function describing the bubble interface is averaged over time and space to generate the void fraction distribution over a slice plane.

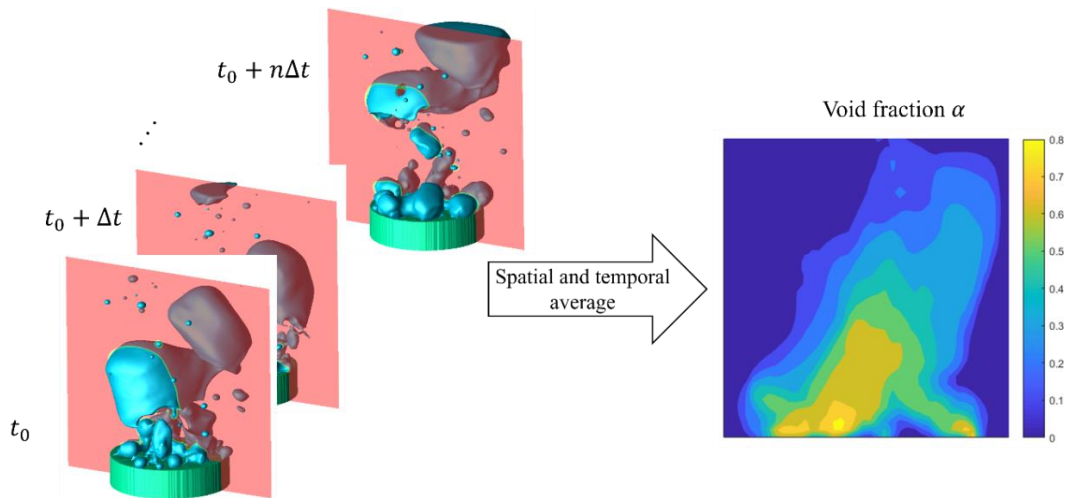

**Figure 3. Demonstration of the average process, from bubble interface to void fraction.**

### 4. Problem setup

In this work, the pool boiling simulation results obtained from Sato et al. [43] is used. The simulation uses color function to resolve the liquid-vapor interface, uses large eddy simulation (LES) for turbulence modeling. The computational domain is 40mm×40mm×38mm, the upper 32mm representing the fluid domain, and the lower 6mm representing the solid domain. In the solid domain, a cylindrical-shaped copper block of diameter 20mm locates at the center, and the surrounding solid material is the thermal insulation, for which the thermal conductivity is defined to be zero. The number of nucleation sites is derived based on the experimental measurement from Gaertner [12]. The location of those nucleation site was prescribed as a *priori*, and randomly distributed on the heating surface, together with a nucleation activation temperature $T_{act}$. The number of cells for the whole domain is 224×224×360. In this work, only the central part of the near wall region, a 10mm×10mm area, is chosen for data extraction in order to minimize the influence of boundary. The computational domain and the data extracted area is depicted in Figure 4.

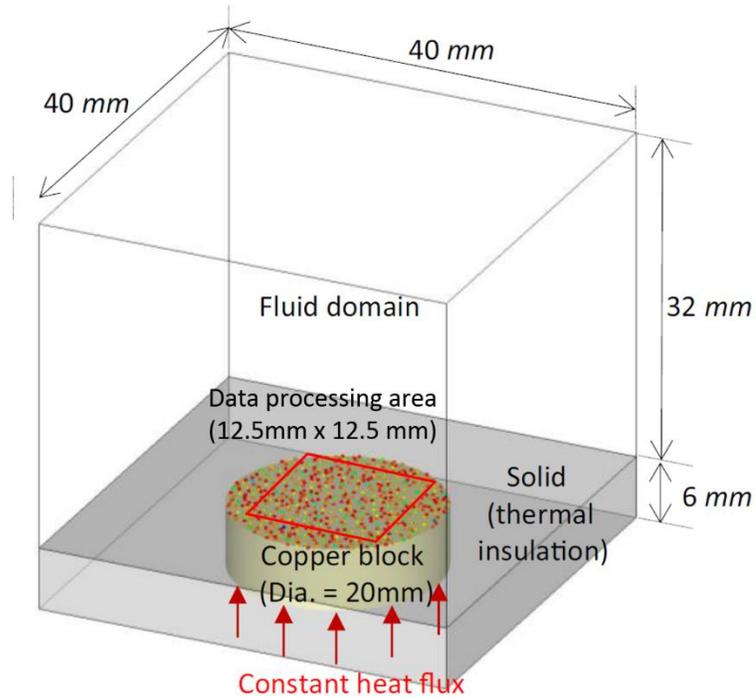

**Figure 4. Computational domain and data sampling area of the ITM simulations.**

The obtained results were validated against experiments from Gaertner[12], and Yabuki and Nakabeppu[53], with good agreement observed on both cases. Thus the simulation can be regarded as a high-fidelity.

The purpose of the desired DFNN is to use local flow features, which can be obtained in the MCFD solver without boiling closure relations, to predict the boiling heat transfer. The input features for DFNN is important and directly influences its performance. In these boiling cases, there is no well-developed bulk flow, thus the widely used dimensionless flow features such as $y^+$ and Reynolds number are not applicable. Considering this, comprehensive flow features are chosen to include all the relevant terms in the averaged conservation equations, including pressure gradient term, momentum convection term, energy convection term, and turbulent viscosity. For simplification purpose, these flow features are further averaged between the vapor phase and liquid phase, thus the number of the input features is significantly reduced. Besides the flow features, the features related to the heating surface, including the total heat flux applied to the heating surface $q_{Total}$, the nucleation site density $N_{site}$, and the nucleation activation temperature $T_{act}$, are also included in the inputs. The selected input features are consistent with the fact that the nucleate boiling is a complex physical process that involves interactions between heating surface, liquid, and vapor. A total 19 features are selected as the inputs of DFNN, which are summarized in Table 1. However, it should be admitted that there should exist more concise input features that represent the physical characteristics of the boiling process. As the first effort to predict the local boiling process with DFNN, the investigation to find better input features is not conducted in this paper.

The wall boiling closure in MCFD provide predictions on heat partitioning and wall superheat. For the prediction of departure from nucleate boiling (DNB) in MCFD solver, the near wall void fraction is also a key parameter. Thus, in this work, these four QoIs are set to be the outputs of the DFNN which are summarized in Table 2.

It is further assumed that boiling is only influenced by the near wall flow based on the scale-separation assumption [48], thus only the near wall flow features are extracted from the ITM results for training the DFNN. In the processing, the average length scale 0.25mm, the average time step is 0.1s. Following the procedure, the near wall region of the whole 224×224×360 computational domain in 100 time frames are averaged to 50×50 near wall flow features, thus 2500 data samples are collected for each simulation case.

**Table 1. Summary of input features of DNN**

| Feature type | Feature expression |
|---|---|
| Pressure gradient | $\frac{\partial \langle p \rangle}{\partial x}$ |
| | $\frac{\partial \langle p \rangle}{\partial y}$ |
| | $\frac{\partial \langle p \rangle}{\partial z}$ |
| Momentum convection | $\frac{\partial \langle \rho \rangle \langle u \rangle \langle u \rangle}{\partial x}$ |
| | $\frac{\partial \langle \rho \rangle \langle u \rangle \langle v \rangle}{\partial x}$ |
| | $\frac{\partial \langle \rho \rangle \langle u \rangle \langle w \rangle}{\partial x}$ |
| | $\frac{\partial \langle \rho \rangle \langle u \rangle \langle v \rangle}{\partial y}$ |
| | $\frac{\partial \langle \rho \rangle \langle v \rangle \langle v \rangle}{\partial y}$ |
| | $\frac{\partial \langle \rho \rangle \langle v \rangle \langle w \rangle}{\partial y}$ |
| | $\frac{\partial \langle \rho \rangle \langle u \rangle \langle w \rangle}{\partial z}$ |
| | $\frac{\partial \langle \rho \rangle \langle v \rangle \langle w \rangle}{\partial z}$ |
| | $\frac{\partial \langle \rho \rangle \langle w \rangle \langle w \rangle}{\partial z}$ |
| Energy convection | $\frac{\partial \langle \rho \rangle \langle h \rangle \langle u \rangle}{\partial x}$ |
| | $\frac{\partial \langle \rho \rangle \langle h \rangle \langle v \rangle}{\partial y}$ |
| | $\frac{\partial \langle \rho \rangle \langle h \rangle \langle w \rangle}{\partial z}$ |
| Turbulence viscosity | $\mu_t$ |
| Heat flux applied to heating surface | $q_{Total}$ |
| Potential nucleation site density | $N_{site}$ |
| Nucleation activation temperature | $T_{act}$ |

**Table 2. Outputs of DFNN**

| QoIs | notation |
|---|---|
| Evaporation heat transfer | $q_{Evap}$ |
| Convective heat transfer towards liquid | $q_{Single}$ |
| Near wall void fraction | $\alpha_{wall}$ |
| Wall superheat | $T_{sup}$ |

For each simulation, constant heat flux is applied at the bottom of the cylindrical copper block. Due to the effect of conjugate heat transfer, randomly distributed nucleation sites, and the random activation temperature assigned to them, the heat flux imposes on the liquid contact surface varies significantly. Thus, the data collect from it covers a broad range of inputs and outputs for the DFNN. Thus, the trained DFNN would have good generalization capability.

In this work, the simulation cases of four different heat flux are used: 600 kW/m², 800 kW/m², 1000 kW/m², and 1200 kW/m². The extracted QoIs from 4 different simulation cases are illustrated in Figure 5 in the form of histogram. As it can be observed, each case has a significantly different distribution, which means that each dataset reflects a different pattern.

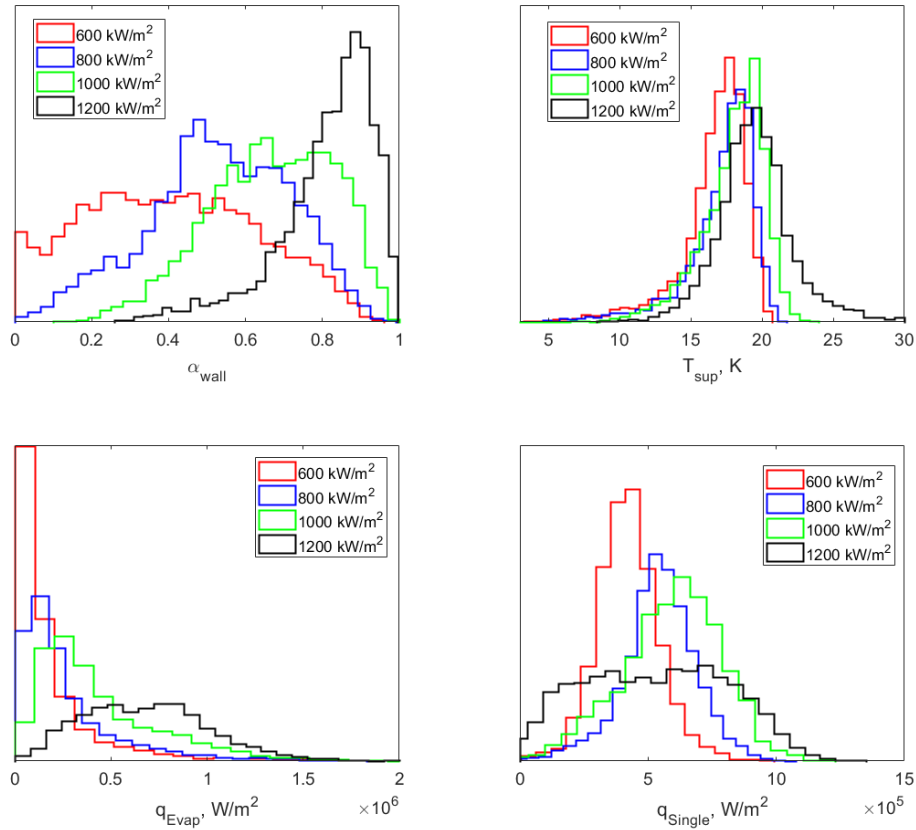

**Figure 5. Histogram of 4 QoIs on different input heat fluxes.**

To summarize, the DFNN is used to predict the near wall boiling behavior, with the averaged near wall flow features that compatible to the two-fluid model as inputs. The high-fidelity simulation results from four pool boiling simulation cases are used for training the DFNN. The architecture of this DFNN is illustrated in Figure 6.

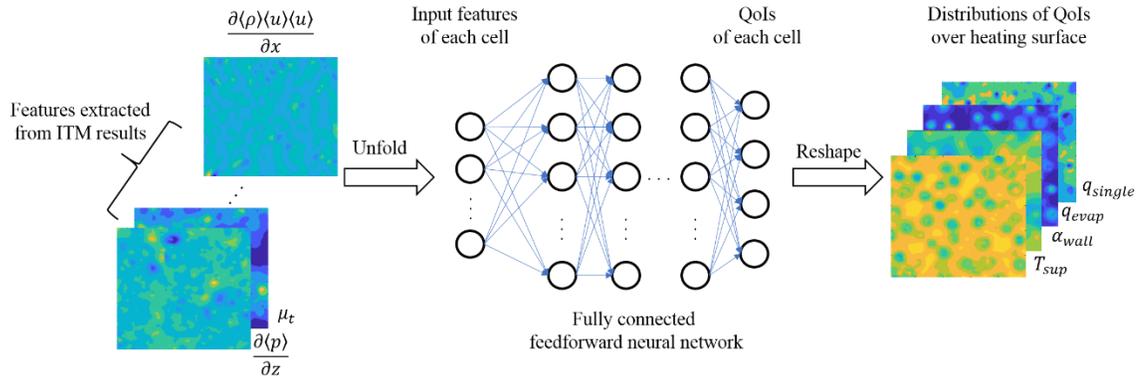

**Figure 6. Architecture of DFNN used for predicting boiling heat transfer.**

## 5. Results and discussion

In this work, the open source deep learning library Pytorch [37] is used for constructing, training and applying the DFNN. All cases are trained on GPU, which is significantly faster compared to training on CPU. To test the performance of the trained network, the full dataset is divided into training dataset and testing dataset. As a rule of thumb, for data of medium size (10000 samples in this work), the training data should be 70%-80% of the full dataset, while the rest for testing. Based on this, four different cases are studied as summarized in Table 3. Each case chose a different simulation results as testing dataset, and three other simulation results as training datasets. Thus, four different DFNNs are trained and tested.

**Table 3. Case studies based on different training/testing data decomposition**

| Cases | Training datasets | Testing dataset |
|---|---|---|
| Case 1 | 800 kW/m$^2$, 1000 kW/m$^2$, 1200 kW/m$^2$ | 600 kW/m$^2$ |
| Case 2 | 600 kW/m$^2$, 1000 kW/m$^2$, 1200 kW/m$^2$ | 800 kW/m$^2$ |
| Case 3 | 600 kW/m$^2$, 800 kW/m$^2$, 1200 kW/m$^2$ | 1000 kW/m$^2$ |
| Case 4 | 600 kW/m$^2$, 800 kW/m$^2$, 1000 kW/m$^2$ | 1200 kW/m$^2$ |

Compared to choosing 25% of the results from each simulation as testing data, this decomposition increased the difficulty of training the DFNN. On the other hand, the DFNN trained in this way should have better generalization capability for unknown inputs, thus can be regarded as better predictive capability. From the perspective of traditional regression, Case 2 and 3 are interpolation cases, while Case 1 and 4 are extrapolation cases. With traditional regression method such as Gaussian process, the interpolation cases should have better accuracy compared to the extrapolation cases.

Before putting into the DFNN, all input features and output QoIs are zero centered and normalized by the mean and standard deviation of the training dataset. As an approach to avoid the vanishing/exploding gradient during the training process, the weights are initialized with the method (known as "Xavier initialization") proposed by Glorot and Bengio[15]. The ELU activation function defined in Section 2 is used. The $L_2$ loss function is used, with $L_2$ regularization term considered. The regularization coefficient λ is tuned to minimize the difference of loss between training data and testing data during the training process. Through a series of testing, λ is set and fixed at 0.0001 for all cases.

It should also be noted that, the influence of hyperparameter is still an active research topic in the deep learning community. The most suitable combination hyperparameters depends on the data and selected input features. Due to the complexity structure of DNN, the theory of optimization of it is still an active research topic. Currently, the choice of hyperparameter is still done through an *ad hoc* analysis. Considering this, the effects of other hyperparameters, including learning rate ε, number of hidden units, and batch size, are also tested based on case 1. The three parameters are sampled over a series of discrete value using the Latin hypercube sampling (LHS) method. A few selected results regarding to the loss on test dataset are demonstrated on Figure 7, in which the average of the root mean square errors (RMSE) of the four QoIs are calculated:

$$RMSE = \sqrt{\frac{1}{N}\sum_{i=1}^{N} L_2(\hat{\boldsymbol{y}}^{(i)}, \boldsymbol{y}^{(i)})} \quad (15)$$

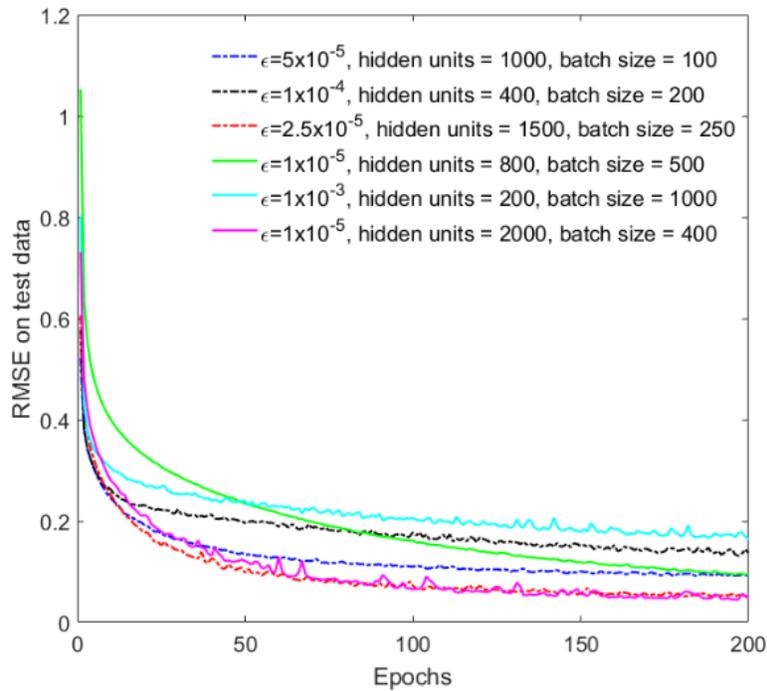

**Figure 7. Demonstrating of hyperparameter influence on DFNN performance.**

As can be observed in Figure 7, all the three hyperparameters have significant influence on the DFNN performance. Several observations can be drawn from this test:

- Too large learning rate $\epsilon$ is a negative factor for the DFNN's performance, whereas a too small value would not help increase the performance but rather slow the training process.
- Generally speaking, a small size of hidden units cannot have good generalization performance on the testing data.
- Large batch size is more computationally efficient on GPU, but cannot produce better performance. In this case, the batch size around 200~500 produce quite similar results.

In this paper, the hyperparameters of the DNNs are tuned and adjust through trial-and-error for each case respectively to achieve minimum error in the testing dataset.

The performance of the trained DFNN can be examined through the scatter plots of DFNN predictions on the testing dataset against the original simulation results. Figure 8 and Figure 9 demonstrate the results of the two interpolation cases. In the figures, the solid $45°$ angle line stands for the ideal situation where DFNN predictions perfectly match the simulation results. The dashed line stands for the $2\sigma$ bound, where $\sigma$ is the standard deviation calculated from the differences between DFNN predictions and the original simulation results. The error bound $2\sigma$ is less than 20% of the averaged QoIs in both cases. Considering the complexity of the problem, and the large uncertainty of the classical wall boiling closures, the prediction of DFNN has lower uncertainty and thus can be regarded as an improvement for this problem. It can be further observed from both figures that although some outliers are present, most of the DFNN predictions are within the $2\sigma$ bound. Considering there are 2500 samples in the plot, those few outliers are not statistically significant.

Moreover, it can be observed in Figure 8 that the DFNN predictions on $\alpha$ are evenly distributed in the $2\sigma$ region, demonstrating a Gaussian distribution pattern, but the other three QoIs show certain skewness. Especially for $q_{Evap}$ and $q_{Single}$, where in general $q_{Evap}$ is underestimated and $q_{Single}$ is overestimated by the DFNN prediction. Similar, but less significant, trends can also be observed in Figure 9. In deep learning community, such discrepancy is termed as high variance (noted that it is a different definition of variance compared to the widely used definition in statistics community) and can usually be resolved by increasing the training data.

The two extrapolation cases are depicted in Figure 10 and Figure 11. Again, the high variance problem is observed in Figure 10, where $T_{sup}$ and $q_{Single}$ are overestimated, whereas $q_{Evap}$ is underestimated. On the other hand, it is also observed that both the $2\sigma$ bound and the scatter plots show quite similar pattern compared to the interpolation case. This indicates the DFNN generalized both the interpolation and the extrapolation cases with similar performance.

A more quantitative comparison, i.e. the RMSE of the DFNNs on the prediction of test datasets are summarized in Table 4. It can be found the RMSE for all QoIs are below 20% of the averaged values. This confirms the DFNNs predict the original simulation results with reasonably good agreement.

On the other hand, it is also noticed that the absolute value of the error is evenly distributed in the whole testing dataset. This means if judging by the error in percentage, the errors for small value predictions would be significantly higher than the large value predictions. This issue stems from the optimization process applied in current DFNN training. In the training process, the whole datasets are divided into several batches, each batch contains hundreds of data. The weights and biases of the DFNN are updated for each batch. This is a much more stable and efficient method compared to update the weights and biases for every single datum. However, when using the $L_2$ norm loss function will average the data in each batch, thus small value data would be less important compared to large value data. In this sense, the trained DFNN would favor the large value data. To overcome this issue, new forms of loss function that specially designed for physical problems should be developed.

Table 4. RMSE on test data of each case

|  | $\alpha_{wall}$ | $T_{sup}$, K | $q_{Evap}$, W/m² | $q_{Single}$, W/m² |
|---|---|---|---|---|
| Case 1 | 0.074 | 1.57 | $7.17 \times 10^4$ | $7.29 \times 10^4$ |
| Case 2 | 0.070 | 1.27 | $8.15 \times 10^4$ | $8.01 \times 10^4$ |
| Case 3 | 0.068 | 1.29 | $9.73 \times 10^4$ | $9.65 \times 10^4$ |
| Case 4 | 0.057 | 1.59 | $1.07 \times 10^5$ | $9.67 \times 10^4$ |

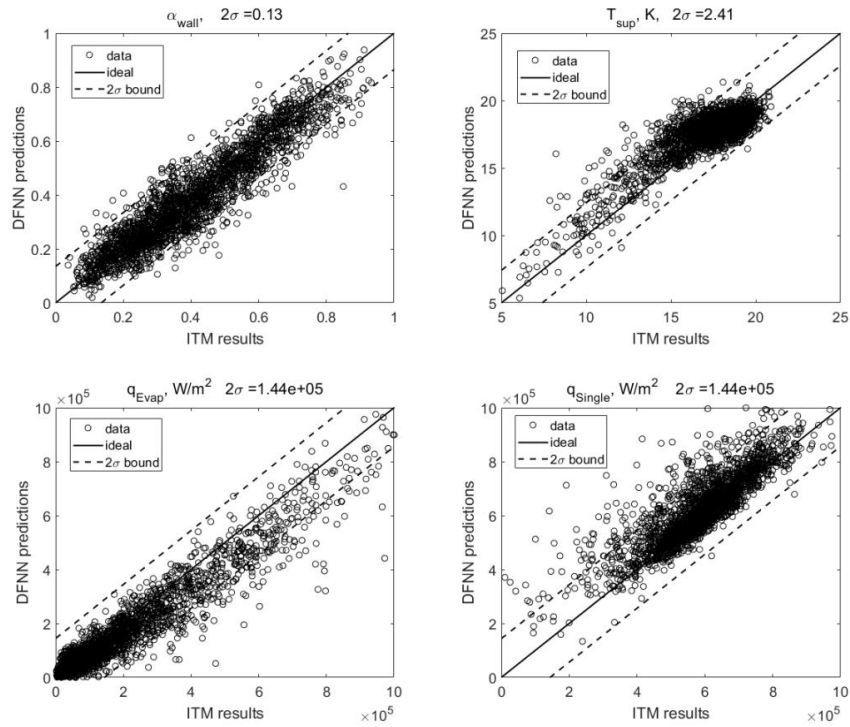

**Figure 8. Comparison between DFNN predictions and real ITM simulations (Case 2).**

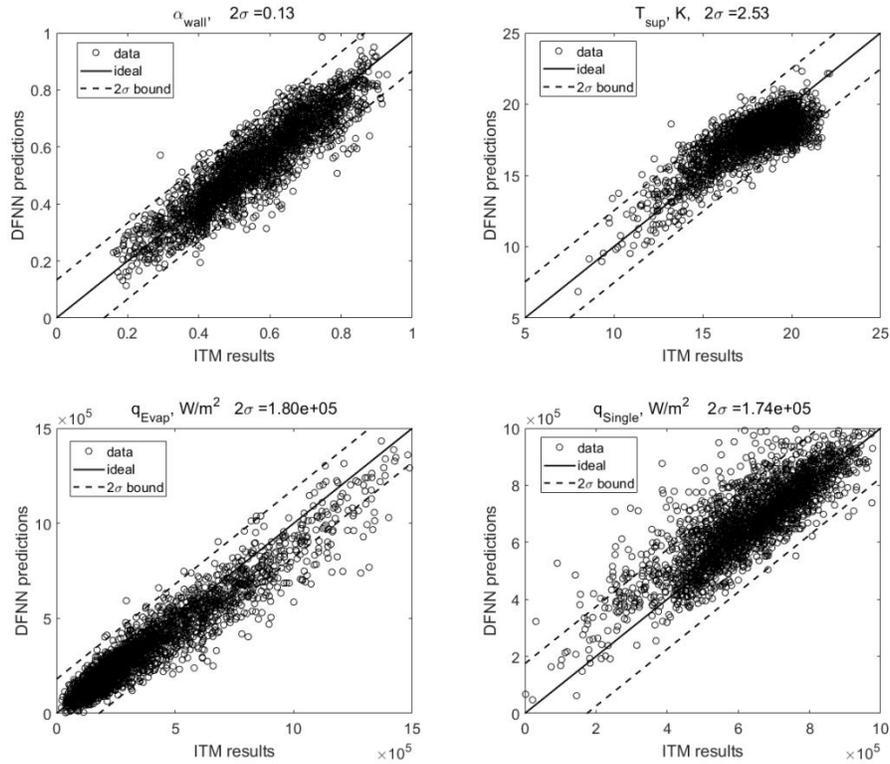

**Figure 9. Comparison between DFNN predictions and real ITM simulations (Case 3).**

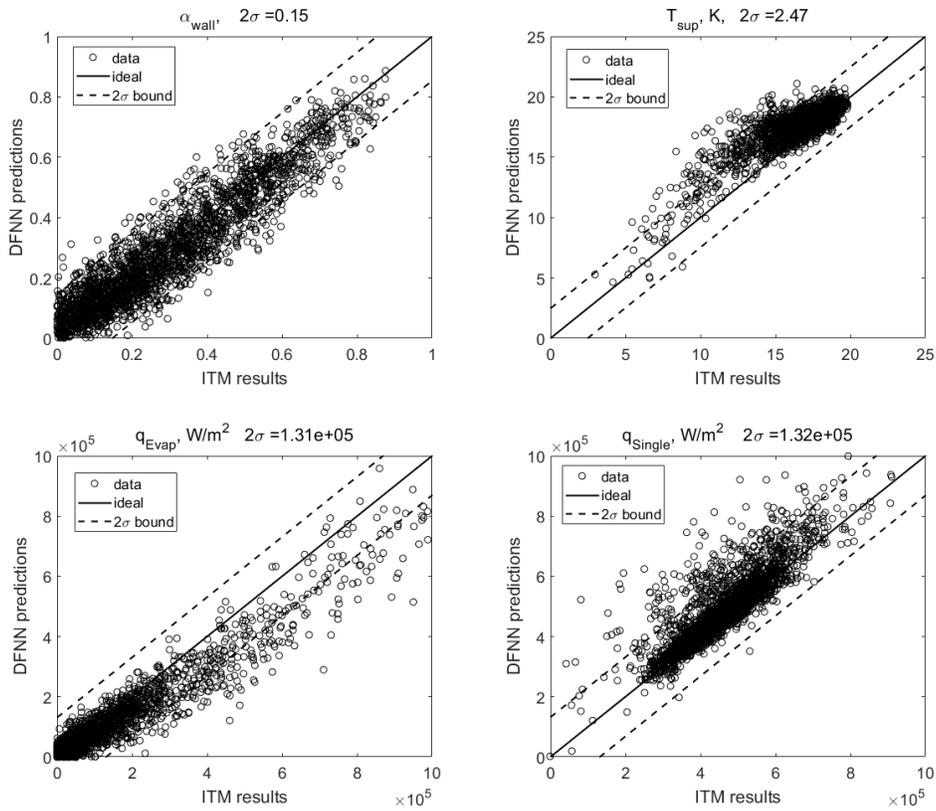

**Figure 10. Comparison between DFNN predictions and real ITM simulations (Case 1).**

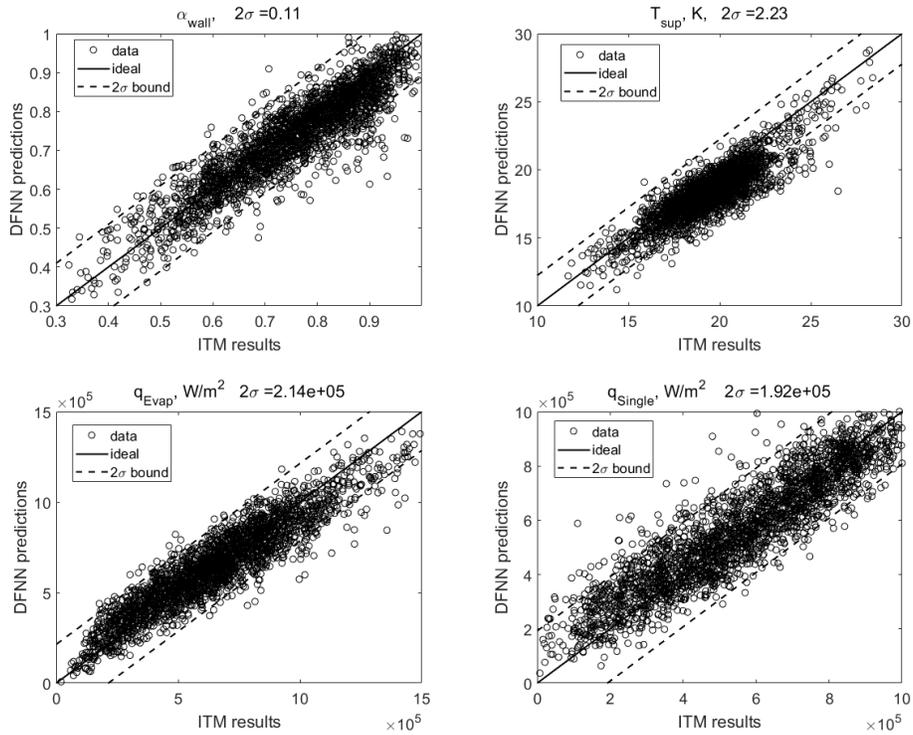

**Figure 11. Comparison between DFNN predictions and real ITM simulations (Case 4).**

The predictive capability of the DFNN can be further demonstrated through the global boiling pattern prediction. The visual comparison between DFNN predictions and the original ITM results on the heating surface is depicted in Figure 12 (Case 2) and Figure 13 (Case 4). The original ITM simulation results suggest two different boiling patterns in these two cases. In Case 2 ($q_{total} = 800 \text{ kW/m}^2$), the individual nucleation sites can be clearly identified which suggested the frequently activated nucleation location in the simulation. The boiling in this case is well-developed nucleate boiling. Whereas in Case 4 ($q_{total} = 1200 \text{ kW/m}^2$), the results show the trend of transition from nucleate boiling to film boiling. For both cases, it can be found that the DFNN prediction captures the global boiling pattern with good accuracy. This suggests that the DFNN relies on local features can not only give good prediction on local boiling process, but also captures the global boiling pattern.

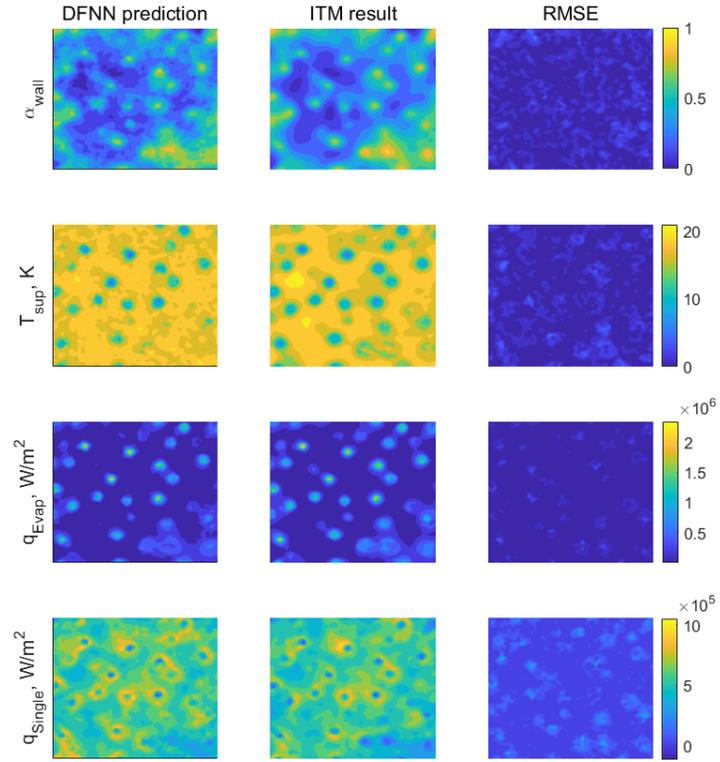

**Figure 12. Visual comparison of DFNN predictions and original ITM resullts (Case 2).**

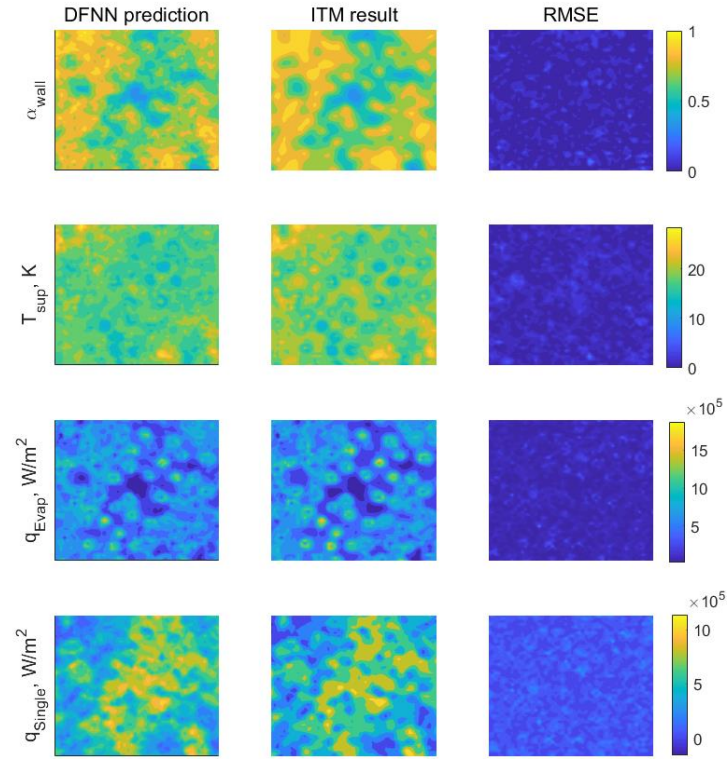

**Figure 13. Visual comparison of DFNN predictions and original ITM resullts (Case 4).**

The mathematical properties of the DNN are still under active investigation, thus the good performances of DFNN on extrapolation prediction observed in this paper still lack solid explanation. Here two possible reasons are developed to explain such good performances. First, the DFNNs developed in this paper rely on the local features extracted from the high-fidelity simulation results. These local features, even under different global conditions (in this case, the different heat fluxes), can still have considerable overlaps, as Figure 5 can serve as an example. In this sense, the global extrapolation is converted to the local interpolation by applying local features to DFNN.

More importantly, these local features preserve certain intrinsic pattern which can be used to characterize the physical process. For example, in fully turbulent flow, the near wall flow can be characterized by the boundary layer theory, no matter the Reynolds number is 50,000 or 80,000. This means the wall function based on boundary layer theory can be extrapolated to predict the near wall flow for any fully turbulent flow case.

On the one hand, traditional regression methods, such as Gaussian process, cannot identify such pattern from the local features (like boundary layer theory), even with unlimited data. The main reason is that these regression methods have limited capability in both expressiveness and generalization. Limited expressiveness capability means these methods cannot approximate complex functional forms if not specified first. Limited generalization capability means these methods cannot identify the intrinsic pattern of a certain problem from a large dataset. In this sense, these traditional regression methods cannot be trusted for extrapolative prediction, as has been already suggested in various practices.

On the other hand, the DNN has demonstrated good capabilities in both expressiveness and generalization. Such good capabilities serve as the second reason for the DFNN on extrapolation prediction. As is discussed in Section 1, a properly trained DNN can approximate any form of continuous function. Also, the DNN has demonstrated good generalization property in the application of natural language processing and computer vision. In this sense, the DNN has the potential to identify the intrinsic pattern of a real physical problem from a large set of data and describe this pattern with enough accuracy. From this point of view, the DNN can serve not only as a statistical tool to replace certain closure relation but also as a promising tool to help discover insights of complex physical problems. The latter depends on the progress of the interpretability of DNN, which means the logic of DNN predictions hidden inside the network should be made understandable by people. This is a very challenging topic but already attracts interests of many researchers [55].

## 6. Conclusion

In this paper, the data-driven approach is studied which takes local flow surface features as inputs to predict the heat transfer behavior in pool boiling using deep feedforward neural networks (DFNN). The networks are trained on data extracted from high-fidelity pool boiling simulations with interface tracking method (ITM). The accuracy of the networks is tested through four case studies, including both interpolation and extrapolation cases. Reasonably good agreement between the DFNN predictions and the original ITM simulations are found for both the interpolation case and the extrapolation case. Moreover, the global boiling pattern over the heating surface can be captured with the DFNN. Especially, the boiling patterns can be captured through DFNN even for extrapolation case. This indicates the deep network trained with local flow features have good generalization property and can thus be trusted to be extended to unknown conditions. The results demonstrate the deep neural networks can be a promising tool to help improve the predictive capability of MCFD solvers. Furthermore, two possible reasons are proposed to explain the good performance of DFNN on the extrapolation prediction. Based on the two reasons, the DNN has the potential to become a promising tool to help discover insights of complex physical problems.

On the other hand, there are still limitations of this work. Firstly, the networks studied in this work are trained with pool boiling simulation results, while for industrial problems, the flow boiling is usually the focus. Due to the significantly different flow and boiling patterns between the two scenarios, the DFNN trained with pool boiling data is questionable to be applied to flow boiling scenarios. Moreover, the flow features chosen in this work are based on the terms of conservation equations, most of which are scale and geometry variant. If a network is designed to have universal predictive capabilities, its features should be scale invariant, as demonstrated in [31]. It worth noting that a most recent work showed promising results to use convolutional neural network to automatically extract physics based features [47]. Finally, the accurate prediction for boiling heat transfer depends on the accurate input of flow features. Such premise cannot be guaranteed with flow features calculated from a MCFD solver. In this sense, to incorporate the deep network in a MCFD solver, the network should also be coupled with that solver in the training process, as is suggested in [5].

## Appendix A. Algorithm of backward propagation

For a DFNN with $l$ layers, $x$ as input features, $y$ the output QoIs, the backward propagation of loss derivatives is applied in following way [16]:

(1). Initialize the weights $\boldsymbol{W}^{(i)}$ and biases $\boldsymbol{b}^{(i)}$ for each layer, $i \in \{1, \ldots, l\}$

(2). Do a forward pass:

$\boldsymbol{h}^{(0)} = \boldsymbol{x}$

FOR k = 1, ..., $l$ do

$\boldsymbol{z}^{(k)} = \boldsymbol{W}^{(k)} \boldsymbol{h}^{(k-1)} + \boldsymbol{b}^{(k)}$
$\boldsymbol{h}^{(k)} = g(\boldsymbol{z}^{(k)})$

END FOR

$\hat{\boldsymbol{y}} = \boldsymbol{h}^{(l)}$

(3). Calculate the loss function, taking the regularization into account:

$J = L(\hat{\boldsymbol{y}}, \boldsymbol{y}) + \lambda \Omega(\boldsymbol{\theta})$

After the forward pass, do the backward propagation of loss derivatives:

(4). Compute the gradient on the output layer:

$\boldsymbol{grad} \leftarrow \nabla_{\hat{\boldsymbol{y}}} J = \nabla_{\hat{\boldsymbol{y}}} L(\hat{\boldsymbol{y}}, \boldsymbol{y})$

(5). Convert the gradient on layer's output into gradient on the pre-nonlinearity activation $\boldsymbol{a}^{(k)}$

FOR k = $l, l-1 \ldots, 1$ do

$\boldsymbol{grad} \leftarrow \nabla_{\boldsymbol{z}^{(k)}} J = \boldsymbol{grad} \odot g'(\boldsymbol{z}^{(k)})$, $\odot$ stands for element-wise multiplication

(6). Compute the gradients on weights and biases:

$\nabla_{\boldsymbol{b}^{(k)}} J = \boldsymbol{grad} + \lambda \nabla_{\boldsymbol{b}^{(k)}} \Omega(\boldsymbol{\theta})$

$\nabla_{\boldsymbol{W}^{(k)}} J = \boldsymbol{grad} \cdot \boldsymbol{h}^{(k-1)T} + \lambda \nabla_{\boldsymbol{W}^{(k)}} \Omega(\boldsymbol{\theta})$

(7). Propagate the gradients to the next lower level hidden layer:

$\boldsymbol{grad} \leftarrow \nabla_{\boldsymbol{h}^{(k-1)}} J = \boldsymbol{W}^{(k)T} \cdot \boldsymbol{grad}$

END FOR

**Nomenclature**

| Physical properties | | | |
|---|---|---|---|
| $C_p$ | specific heat at constant pressure, J/(kg·K) | *Greek symbols* | |
| $\boldsymbol{f}$ | body force density, N/m$^3$ | $\alpha$ | void fraction |
| $h$ | specific enthalpy, J/kg | $\Gamma_{lv}$ | evaporation rate, kg/(m$^3$·s) |
| $N$ | nucleation site density, 1/m$^2$ | $\lambda$ | thermal conductivity, W/(m·K) |
| $p$ | pressure, Pa | $\mu$ | dynamic viscosity, Pa·s |
| $q$ | heat flux, W/m$^2$ | $\rho$ | density, kg/m$^3$ |
| $Q$ | heat source, W/m$^3$ | $\phi$ | color function |
| $T$ | temperature, K | | |
| $t$ | time, s | | |
| $\boldsymbol{U}$ | velocity vector, m/s | *subscripts* | |
| $u$ | velocity component on $x$ direction, m/s | $l$ | liquid phase |
| $v$ | velocity component on $y$ direction, m/s | $t$ | turbulent |
| $w$ | velocity component on $z$ direction, m/s | $v$ | vapor phase |
| | | | |
| Neural network parameters | | | |
| $\boldsymbol{b}$ | bias | *Greek symbols* | |
| $g$ | nonlinear activation function | $\varepsilon$ | learning rate |
| $h$ | output of hidden layer | $\lambda$ | regularization coefficient |
| $J$ | total loss function | $\boldsymbol{\theta}$ | learnable parameters |
| $L$ | loss function | $\Omega$ | regularization term |
| $\boldsymbol{w}$ | weight vector | | |
| $\boldsymbol{W}$ | weight matrix | | |
| $\boldsymbol{x}$ | input feature | | |
| $\boldsymbol{y}$ | output | | |
| $\boldsymbol{z}$ | pre-nonlinearity activation | | |

**Acknowledgement**

This research was partially supported by the Consortium for Advanced Simulation of Light Water Reactors (http://www.casl.gov), an Energy Innovation Hub (http://www.energy.gov/hubs) for Modeling and Simulation of Nuclear Reactors under U.S. Department of Energy Contract No.

DE-AC05-00OR22725. and by Nuclear Energy University Program under the grant DE-NE0008530.